

\font \eightbf         = cmbx8
\font \eighti          = cmmi8
\font \eightit         = cmti8
\font \eightrm         = cmr8
\font \eightsl         = cmsl8
\font \eightsy         = cmsy8
\font \eighttt         = cmtt8
\font \tenbf           = cmbx9
\font \teni            = cmmi9
\font \tenit           = cmti9
\font \tenrm           = cmr9
\font \tensl           = cmsl9
\font \tensy           = cmsy9
\font \tentt           = cmtt9

\font \kleinhalbcurs   = cmmib10 scaled 800

\font \sixbf           = cmbx6
\font \sixi            = cmmi6
\font \sixrm           = cmr6
\font \sixsy           = cmsy6
\font \tafonts         = cmbx12
\font \tafontss        = cmbx10
\font \tafontt         = cmbx10 scaled\magstep2
\font \tams            = cmmib10
\font \tenmib          = cmmib10
\font \tamt            = cmmib10
\font \tass            = cmsy10
\font \tasss           = cmsy7
\font \tast            = cmsy10 scaled\magstep2
\font \tbfonts         = cmbx8
\font \tbfontss        = cmbx10  scaled 667
\font \tbfontt         = cmbx10 scaled\magstep1
\font \tbmt            = cmmib10
\font \tbss            = cmsy8
\font \tbsss           = cmsy6
\font \tbst            = cmsy10  scaled\magstep1
\vsize=23.5truecm
\hoffset=-1true cm
\voffset=-1true cm
\newdimen\fullhsize
\fullhsize=40cc
\hsize=19.5cc
\def\fullline{\hbox to\fullhsize}
\def\makefootline{\baselineskip=10dd \fullline{\the\footline}}
\def\makeheadline{\vbox to 0pt{\vskip-22.5pt
            \fullline{\vbox to 8.5pt{}\the\headline}\vss}\nointerlineskip}
\let\lr=L \newbox\leftcolumn
\output={\global\topskip=10pt
         \if L\lr
            \global\setbox\leftcolumn=\columnbox \global\let
\lr=R
            \message{[left\the\pageno]}%
            \ifnum\pageno=1
               \global\topskip=\fullhead\fi
         \else
            \doubleformat \global\let\lr=L
         \fi
         \ifnum\outputpenalty>-2000 \else\dosupereject\fi}
\def\doubleformat{\shipout\vbox{\makeheadline
    \fullline{\box\leftcolumn\hfil\columnbox}
           \makefootline}
           \advancepageno}
\def\columnbox{\leftline{\pagebody}}
\outer\def\bye{\bigskip\typeset
\sterne=1\ifx\speciali\undefined\else
\loop\smallskip\noindent special character No\number\sterne:
\csname special\romannumeral\sterne\endcsname
\advance\sterne by 1\global\sterne=\sterne
\ifnum\sterne<11\repeat\fi
\if R\lr\null\fi\vfill\supereject\end}
\def\typeset{\begpet\noindent This article was processed by the author
 using
Sprin\-ger-Ver\-lag \TeX\ AA macro package 1989.\endpet}
\hfuzz=2pt
\vfuzz=2pt
\tolerance=1000
\fontdimen3\tenrm=1.5\fontdimen3\tenrm
\fontdimen7\tenrm=1.5\fontdimen7\tenrm
\abovedisplayskip=3 mm plus6pt minus 4pt
\belowdisplayskip=3 mm plus6pt minus 4pt
\abovedisplayshortskip=0mm plus6pt
\belowdisplayshortskip=2 mm plus4pt minus 4pt
\predisplaypenalty=0
\clubpenalty=20000
\widowpenalty=20000
\parindent=1.5em
\frenchspacing
\def\newline{\hfill\break}%
\nopagenumbers
\def\AALogo{\setbox254=\hbox{ ASTROPHYSICS }%
\vbox{\baselineskip=10dd\hrule\hbox{\vrule\vbox{\kern3pt
\hbox to\wd254{\hfil ASTRONOMY\hfil}
\hbox to\wd254{\hfil AND\hfil}\copy254
\hbox to\wd254{\hfil\number\day.\number\month.\number\year\hfil}
\kern3pt}\vrule}\hrule}}
\def\paglay{\headline={{\tenrm\hsize=.75\fullhsize\ifnum\pageno=1
\vbox{\baselineskip=10dd\hrule\line{\vrule\kern3pt\vbox{\kern3pt
\hbox{\bf A and A Manuskript-Nr.}
\hbox{(will be inserted by hand later)}
\kern3pt\hrule\kern3pt
\hbox{\bf Your thesaurus codes are:}
\hbox{\rightskip=0pt plus3em\advance\hsize by-7pt
\vbox{\noindent\ignorespaces\the\THESAURUS}}
\kern3pt}\hfil\kern3pt\vrule}\hrule}
\rlap{\quad\AALogo}\hfil
\else\ifodd\pageno\hfil\folio\else\folio\hfil\fi\fi}}}
\ifx \undefined\instruct
\headline={\tenrm\ifodd\pageno\hfil\folio
\else\folio\hfil\fi}\fi
\newcount\eqnum\eqnum=0%
\def\autnum{\global\advance\eqnum by 1{\rm(\the\eqnum)}}
\newtoks\eq\newtoks\eqn
\catcode`@=11
\def\eqalign#1{\null\vcenter{\openup\jot\m@th
  \ialign{\strut\hfil$\displaystyle{##}$&$\displaystyle{{}##}$
\hfil
      \crcr#1\crcr}}}
\def\displaylines#1{{}$\displ@y
\hbox{\vbox{\halign{$\@lign\hfil\displaystyle##\hfil$\crcr
    #1\crcr}}}${}}
\def\eqalignno#1{{}$\displ@y
  \hbox{\vbox{\halign to\hsize{\hfil$\@lign\displaystyle{##}$\tabskip
\z@skip
    &$\@lign\displaystyle{{}##}$\hfil\tabskip\centering
    &\llap{$\@lign##$}\tabskip\z@skip\crcr
    #1\crcr}}}${}}
\def\leqalignno#1{{}$\displ@y
\hbox{\vbox{\halign to\hsize{\qquad\hfil$\@lign\displaystyle{##}$\tabskip
\z@skip
    &$\@lign\displaystyle{{}##}$\hfil\tabskip\centering
    &\kern-\hsize\rlap{$\@lign##$}\tabskip\hsize\crcr
    #1\crcr}}}${}}
\def\generaldisplay{%
\ifeqno
       \ifleqno\leftline{$\displaystyle\the\eqn\quad\the\eq$}%
       \else\line{$\displaystyle\the\eq\hfill\the\eqn$}\fi
\else
       \leftline{$\displaystyle\the\eq$}%
\fi
\global\eq={}\global\eqn={}}%
\newif\ifeqno\newif\ifleqno \everydisplay{\displaysetup}
\def\displaysetup#1$${\displaytest#1\eqno\eqno\displaytest}
\def\displaytest#1\eqno#2\eqno#3\displaytest{%
\if!#3!\ldisplaytest#1\leqno\leqno\ldisplaytest
\else\eqnotrue\leqnofalse\eqn={#2}\eq={#1}\fi
\generaldisplay$$}
\def\ldisplaytest#1\leqno#2\leqno#3\ldisplaytest{\eq={#1}%
\if!#3!\eqnofalse\else\eqnotrue\leqnotrue\eqn={#2}\fi}
\catcode`@=12 %
\mathchardef\Gamma="0100
\mathchardef\Delta="0101
\mathchardef\Theta="0102
\mathchardef\Lambda="0103
\mathchardef\Xi="0104
\mathchardef\Pi="0105
\mathchardef\Sigma="0106
\mathchardef\Upsilon="0107
\mathchardef\Phi="0108
\mathchardef\Psi="0109
\mathchardef\Omega="010A

\def\utw{\smash{\rlap{\lower5pt\hbox{$\sim$}}}}
\def\udtw{\smash{\rlap{\lower6pt\hbox{$\approx$}}}}

\def\diameter{{\ifmmode\mathchoice
{\ooalign{\hfil\hbox{$\displaystyle/$}\hfil\crcr
{\hbox{$\displaystyle\mathchar"20D$}}}}
{\ooalign{\hfil\hbox{$\textstyle/$}\hfil\crcr
{\hbox{$\textstyle\mathchar"20D$}}}}
{\ooalign{\hfil\hbox{$\scriptstyle/$}\hfil\crcr
{\hbox{$\scriptstyle\mathchar"20D$}}}}
{\ooalign{\hfil\hbox{$\scriptscriptstyle/$}\hfil\crcr
{\hbox{$\scriptscriptstyle\mathchar"20D$}}}}
\else{\ooalign{\hfil/\hfil\crcr\mathhexbox20D}}%
\fi}}

\normallineskip=1dd
\normallineskiplimit=0dd
\normalbaselineskip=10dd
\textfont0=\tenrm
\textfont1=\teni
\textfont2=\tensy
\textfont\itfam=\tenit
\textfont\slfam=\tensl
\textfont\ttfam=\tentt
\textfont\bffam=\tenbf
\normalbaselines\rm
\def\petit{\def\rm{\fam0\eightrm}%
\textfont0=\eightrm \scriptfont0=\sixrm \scriptscriptfont0=\fiverm
 \textfont1=\eighti \scriptfont1=\sixi \scriptscriptfont1=\fivei
 \textfont2=\eightsy \scriptfont2=\sixsy \scriptscriptfont2=
\fivesy
 \def\it{\fam\itfam\eightit}%
 \textfont\itfam=\eightit
 \def\sl{\fam\slfam\eightsl}%
 \textfont\slfam=\eightsl
 \def\bf{\fam\bffam\eightbf}%
 \textfont\bffam=\eightbf \scriptfont\bffam=\sixbf
 \scriptscriptfont\bffam=\fivebf
 \def\tt{\fam\ttfam\eighttt}%
 \textfont\ttfam=\eighttt
 \let\tams=\kleinhalbcurs
 \let\tenbf=\eightbf
 \let\sevenbf=\sixbf
 \normalbaselineskip=9dd
 \if Y\REFEREE \normalbaselineskip=2\normalbaselineskip
 \normallineskip=2\normallineskip\fi
 \setbox\strutbox=\hbox{\vrule height7pt depth2pt width0pt}%
 \normalbaselines\rm}%
\def\begpet{\vskip6pt\bgroup\petit}%
\def\endpet{\vskip6pt\egroup}%
\def\rahmen#1{\vbox{\hrule\line{\vrule\vbox to#1true
cm{\vfil}\hfil\vrule}\vfil\hrule}}
\def\begfig#1cm#2\endfig{\par
   \ifvoid\topins\midinsert\bigskip\vbox{\rahmen{#1}#2}\endinsert
   \else\topinsert\vbox{\rahmen{#1}#2}\endinsert%
\fi}
\def\begfigwid#1cm#2\endfig{\par
\if N\lr\else
\if R\lr
\shipout\vbox{\makeheadline
\line{\box\leftcolumn}\makefootline}\advancepageno
\fi\let\lr=N
\topskip=10pt
\output={\plainoutput}%
\fi
\topinsert\line{\vbox{\hsize=\fullhsize\rahmen{#1}#2}\hss}\endinsert}

\def\begtab#1cm#2\endtab{\par
   \ifvoid\topins\midinsert\medskip\vbox{#2\rahmen{#1}}\endinsert
   \else\topinsert\vbox{#2\rahmen{#1}}\endinsert%
\fi}
\def\begtabemptywid#1cm#2\endtab{\par
\if N\lr\else
\if R\lr
\shipout\vbox{\makeheadline
\line{\box\leftcolumn}\makefootline}\advancepageno
\fi\let\lr=N
\topskip=10pt
\output={\plainoutput}%
\fi
\topinsert\line{\vbox{\hsize=\fullhsize#2\rahmen{#1}}\hss}\endinsert}
\def\begtabfullwid#1\endtab{\par
\if N\lr\else
\if R\lr
\shipout\vbox{\makeheadline
\line{\box\leftcolumn}\makefootline}\advancepageno
\fi\let\lr=N
\output={\plainoutput}%
\fi
\topinsert\line{\vbox{\hsize=\fullhsize\noindent#1}\hss}\endinsert}

\def\begref{\vskip1cm\begingroup\let\INS=N}
\def\ref{\goodbreak\if N\INS\let\INS=Y\vbox{\noindent\tenbf
References\bigskip}\fi\hangindent\parindent
\hangafter=1\noindent\ignorespaces}
\def\endref{\goodbreak\endgroup}%
\def\ack#1{\vskip11pt\begingroup\noindent{\it Acknowledgements
\/}.
\ignorespaces#1\vskip6pt\endgroup}

 \def \aTa  { \goodbreak
     \bgroup
     \par
 \textfont0=\tafontt \scriptfont0=\tafonts \scriptscriptfont0=
\tafontss
 \textfont1=\tamt \scriptfont1=\tbmt \scriptscriptfont1=\tams
 \textfont2=\tast \scriptfont2=\tass \scriptscriptfont2=\tasss
     \baselineskip=17dd
     \lineskip=17dd
     \rightskip=0pt plus2cm\spaceskip=.3333em \xspaceskip=.5em
     \pretolerance=10000
     \noindent
     \tafontt}
 \def \eTa{\vskip10pt\egroup
     \noindent
     \ignorespaces}

 \def \aTb{\goodbreak
     \bgroup
     \par
 \textfont0=\tbfontt \scriptfont0=\tbfonts \scriptscriptfont0=
\tbfontss
 \textfont1=\tbmt \scriptfont1=\tenmib \scriptscriptfont1=\tams
 \textfont2=\tbst \scriptfont2=\tbss \scriptscriptfont2=\tbsss
     \baselineskip=13dd
     \lineskip=13dd
     \rightskip=0pt plus2cm\spaceskip=.3333em \xspaceskip=.5em
     \pretolerance=10000
     \noindent
     \tbfontt}
 \def \eTb{\vskip10pt
     \egroup
     \noindent
     \ignorespaces}
\catcode`\@=11
\expandafter \newcount \csname c@Tl\endcsname
    \csname c@Tl\endcsname=0
\expandafter \newcount \csname c@Tm\endcsname
    \csname c@Tm\endcsname=0
\expandafter \newcount \csname c@Tn\endcsname
    \csname c@Tn\endcsname=0
\expandafter \newcount \csname c@To\endcsname
    \csname c@To\endcsname=0
\expandafter \newcount \csname c@Tp\endcsname
    \csname c@Tp\endcsname=0
\def \resetcount#1    {\global
    \csname c@#1\endcsname=0}
\def\@nameuse#1{\csname #1\endcsname}
\def\arabic#1{\@arabic{\@nameuse{c@#1}}}
\def\@arabic#1{\ifnum #1>0 \number #1\fi}
\expandafter \newcount \csname c@fn\endcsname
    \csname c@fn\endcsname=0
\def \stepc#1    {\global
    \expandafter
    \advance
    \csname c@#1\endcsname by 1}
\catcode`\@=12
   \catcode`\@= 11

\skewchar\eighti='177 \skewchar\sixi='177
\skewchar\eightsy='60 \skewchar\sixsy='60
\hyphenchar\eighttt=-1
\def\footnoterule{\kern-3pt\hrule width 2true cm\kern2.6pt}%
\newinsert\footins
\def\footnotea#1{\let\@sf\empty %
  \ifhmode\edef\@sf{\spacefactor\the\spacefactor}\/\fi
  {#1}\@sf\vfootnote{#1}}
\def\vfootnote#1{\insert\footins\bgroup
  \textfont0=\tenrm\scriptfont0=\sevenrm\scriptscriptfont0=\fiverm
  \textfont1=\teni\scriptfont1=\seveni\scriptscriptfont1=\fivei
  \textfont2=\tensy\scriptfont2=\sevensy\scriptscriptfont2=\fivesy
  \interlinepenalty\interfootnotelinepenalty
  \splittopskip\ht\strutbox %
  \splitmaxdepth\dp\strutbox \floatingpenalty\@MM
  \leftskip\z@skip \rightskip\z@skip \spaceskip\z@skip \xspaceskip
\z@skip
  \textindent{#1}\footstrut\futurelet\next\fo@t}
\def\fo@t{\ifcat\bgroup\noexpand\next \let\next\f@@t
  \else\let\next\f@t\fi \next}
\def\f@@t{\bgroup\aftergroup\@foot\let\next}
\def\f@t#1{#1\@foot}
\def\@foot{\strut\egroup}
\def\footstrut{\vbox to\splittopskip{}}
\skip\footins=\bigskipamount %
\count\footins=1000 %
\dimen\footins=8in %
   \def \bfootax  {\bgroup\tenrm
                  \baselineskip=12pt\lineskiplimit=-6pt
                  \hsize=19.5cc
                  \def\textindent##1{\hang\noindent\hbox
                  to\parindent{##1\hss}\ignorespaces}%
                  \footnotea{$^\star$}\bgroup}
   \def \efootax  {\egroup\egroup}
   \def \bfootay  {\bgroup\tenrm
                  \baselineskip=12pt\lineskiplimit=-6pt
                  \hsize=19.5cc
                  \def\textindent##1{\hang\noindent\hbox
                  to\parindent{##1\hss}\ignorespaces}%
                  \footnotea{$^{\star\star}$}\bgroup}
   \def \efootay  {\egroup\egroup }
   \def \bfootaz {\bgroup\tenrm
                  \baselineskip=12pt\lineskiplimit=-6pt
                  \hsize=19.5cc
                  \def\textindent##1{\hang\noindent\hbox
                  to\parindent{##1\hss}\ignorespaces}%
                 \footnotea{$^{\star\star\star}$}\bgroup}
   \def \efootaz {\egroup \egroup}
\def\fonote#1{\mehrsterne$^{\the\sterne}$\begingroup
       \def\textindent##1{\hang\noindent\hbox
       to\parindent{##1\hss}\ignorespaces}%
\vfootnote{$^{\the\sterne}$}{#1}\endgroup}
\catcode`\@=12
\everypar={\let\lasttitle=N\everypar={\parindent=1.5em}}%
\def \titlea#1{\stepc{Tl}
     \resetcount{Tm}
     \vskip22pt
     \setbox0=\vbox{\vskip 22pt\noindent
     \bf
     \rightskip 0pt plus4em
     \pretolerance=20000
     \arabic{Tl}.\
    \textfont1=\tams\scriptfont1=\kleinhalbcurs\scriptscriptfont1=
\kleinhalbcurs
     \ignorespaces#1
     \vskip11pt}
     \dimen0=\ht0\advance\dimen0 by\dp0\advance\dimen0 by 2\baselineskip
     \advance\dimen0 by\pagetotal
     \ifdim\dimen0>\pagegoal\eject\fi
     \bgroup
     \noindent
     \bf
     \rightskip 0pt plus4em
     \pretolerance=20000
     \arabic{Tl}.\
    \textfont1=\tams\scriptfont1=\kleinhalbcurs\scriptscriptfont1=
\kleinhalbcurs
     \ignorespaces#1
     \vskip11pt
     \egroup
     \nobreak
     \parindent=0pt
     \everypar={\parindent=1.5em
     \let\lasttitle=N\everypar={\let\lasttitle=N}}%
     \let\lasttitle=A%
     \ignorespaces}
 \def\titleb#1{\stepc{Tm}
     \resetcount{Tn}
     \if N\lasttitle\else\vskip-11pt\vskip-\baselineskip
     \fi
     \vskip 17pt
     \setbox0=\vbox{\vskip 17pt
     \raggedright
     \pretolerance=10000
     \noindent
     \it
     \arabic{Tl}.\arabic{Tm}.\
     \ignorespaces#1
     \vskip8pt}
     \dimen0=\ht0\advance\dimen0 by\dp0\advance\dimen0 by 2\baselineskip
     \advance\dimen0 by\pagetotal
     \ifdim\dimen0>\pagegoal\eject\fi
     \bgroup
     \raggedright
     \pretolerance=10000
     \noindent
     \it
     \arabic{Tl}.\arabic{Tm}.\
     \ignorespaces#1
     \vskip8pt
     \egroup
     \nobreak
     \let\lasttitle=B%
     \parindent=0pt
     \everypar={\parindent=1.5em
     \let\lasttitle=N\everypar={\let\lasttitle=N}}%
     \ignorespaces}
 \def \titlec#1{\stepc{Tn}
     \resetcount{To}
     \if N\lasttitle\else\vskip-3pt\vskip-\baselineskip
     \fi
     \vskip 11pt
     \setbox0=\vbox{\vskip 11pt
     \noindent
     \raggedright
     \pretolerance=10000
     \arabic{Tl}.\arabic{Tm}.\arabic{Tn}.\
     \ignorespaces#1\vskip6pt}
     \dimen0=\ht0\advance\dimen0 by\dp0\advance\dimen0 by 2\baselineskip
     \advance\dimen0 by\pagetotal
     \ifdim\dimen0>\pagegoal\eject\fi
     \bgroup\noindent
     \raggedright
     \pretolerance=10000
     \arabic{Tl}.\arabic{Tm}.\arabic{Tn}.\
     \ignorespaces#1\vskip6pt
     \egroup
     \nobreak
     \let\lasttitle=C%
     \parindent=0pt
     \everypar={\parindent=1.5em
     \let\lasttitle=N\everypar={\let\lasttitle=N}}%
     \ignorespaces}
 \def\titled#1{\stepc{To}
     \resetcount{Tp}
     \if N\lasttitle\else\vskip-3pt\vskip-\baselineskip
     \fi
     \vskip 11pt
     \bgroup
     \it
     \noindent
     \ignorespaces#1\unskip. \egroup\ignorespaces}
\let\REFEREE=N
\newbox\refereebox
\setbox\refereebox=\vbox
to0pt{\vskip0.5cm\fullline{\hrulefill\tentt\lower0.5ex
\hbox{\kern5pt referee's copy\kern5pt}\hrulefill}\vss}%
\def\refereelayout{\let\REFEREE=M\footline={\copy\refereebox}%
\message{|A referee's copy will be produced}\par
\if N\lr\else
\if R\lr
\shipout\vbox{\makeheadline
\line{\box\leftcolumn}\makefootline}\advancepageno
\fi\let\lr=N
\topskip=10pt
\output={\plainoutput}%
\fi
}
\let\ts=\thinspace
\newcount\sterne \sterne=0
\newdimen\fullhead
\newtoks\RECDATE
\newtoks\ACCDATE
\newtoks\MAINTITLE
\newtoks\SUBTITLE
\newtoks\AUTHOR
\newtoks\INSTITUTE
\newtoks\SUMMARY
\newtoks\KEYWORDS
\newtoks\THESAURUS
\newtoks\SENDOFF
\newlinechar=`\| %
\catcode`\@=\active
\let\INS=N%
\def@#1{\if N\INS $^{#1}$\else\if
E\INS\hangindent0.5\parindent\hangafter=1%
\noindent\hbox to0.5\parindent{$^{#1}$\hfil}\let\INS=Y\ignorespaces
\else\par\hangindent0.5\parindent\hangafter=1
\noindent\hbox to0.5\parindent{$^{#1}$\hfil}\ignorespaces\fi
\fi}%
\def\mehrsterne{\advance\sterne by1\global\sterne=\sterne}%
\def\FOOTNOTE#1{\mehrsterne\ifcase\sterne
\or\bfootax \ignorespaces #1\efootax
\or\bfootay \ignorespaces #1\efootay
\or\bfootaz \ignorespaces #1\efootaz\else\fi}%
\def\PRESADD#1{\mehrsterne\ifcase\sterne
\or\bfootax Present address: #1\efootax
\or\bfootay Present address: #1\efootay
\or\bfootaz Present address: #1\efootaz\else\fi}%
\def\maketitle{\paglay%
\def\missing{ ????? }%
\setbox0=\vbox{\parskip=0pt\hsize=\fullhsize\null\vskip2truecm
\let\kka = \tamt
\edef\test{\the\MAINTITLE}%
\ifx\test\missing\MAINTITLE={MAINTITLE should be given}\fi
\aTa\ignorespaces\the\MAINTITLE\eTa
\let\kka = \tbmt
\edef\test{\the\SUBTITLE}%
\ifx\test\missing\else\aTb\ignorespaces\the\SUBTITLE\eTb\fi
\let\kka = \tams
\edef\test{\the\AUTHOR}%
\ifx\test\missing
\AUTHOR={Name(s) and initial(s) of author(s) should be given}
\fi
\noindent{\bf\ignorespaces\the\AUTHOR\vskip4pt}
\let\INS=E%
\edef\test{\the\INSTITUTE}%
\ifx\test\missing
\INSTITUTE={Address(es) of author(s) should be given.}\fi
{\noindent\ignorespaces\the\INSTITUTE\vskip10pt}%
\edef\test{\the\RECDATE}%
\ifx\test\missing
\RECDATE={{\petit $[$the date should be inserted later$]$}}\fi
\edef\test{\the\ACCDATE}%
\ifx\test\missing
\ACCDATE={{\petit $[$the date should be inserted later$]$}}\fi
{\noindent Received \ignorespaces\the\RECDATE\unskip; accepted
\ignorespaces
\the\ACCDATE\vskip21pt\bf S}}%
\global\fullhead=\ht0\global\advance\fullhead by\dp0
\global\advance\fullhead by10pt\global\sterne=0
{\parskip=0pt\hsize=19.5cc\null\vskip2truecm
\edef\test{\the\SENDOFF}%
\ifx\test\missing\else\insert\footins{\smallskip\noindent
{\it Send offprint requests to\/}: \ignorespaces\the\SENDOFF}
\fi
\hsize=\fullhsize
\let\kka = \tamt
\edef\test{\the\MAINTITLE}%
\ifx\test\missing\message{|Your MAINTITLE is missing.}%
\MAINTITLE={MAINTITLE should be given}\fi
\aTa\ignorespaces\the\MAINTITLE\eTa
\let\kka = \tbmt
\edef\test{\the\SUBTITLE}%
\ifx\test\missing\message{|The SUBTITLE is optional.}%
\else\aTb\ignorespaces\the\SUBTITLE\eTb\fi
\let\kka = \tams
\edef\test{\the\AUTHOR}%
\ifx\test\missing\message{|Name(s) and initial(s) of author(s)
 missing.}%
\AUTHOR={Name(s) and initial(s) of author(s) should be given}
\fi
\noindent{\bf\ignorespaces\the\AUTHOR\vskip4pt}
\let\INS=E%
\edef\test{\the\INSTITUTE}%
\ifx\test\missing\message{|Address(es) of author(s) missing.}%
\INSTITUTE={Address(es) of author(s) should be given.}\fi
{\noindent\ignorespaces\the\INSTITUTE\vskip10pt}%
\edef\test{\the\RECDATE}%
\ifx\test\missing\message{|The date of receipt should be inserted
later.}%
\RECDATE={{\petit $[$the date should be inserted later$]$}}\fi
\edef\test{\the\ACCDATE}%
\ifx\test\missing\message{|The date of acceptance should be inserted
later.}%
\ACCDATE={{\petit $[$the date should be inserted later$]$}}\fi
{\noindent Received \ignorespaces\the\RECDATE\unskip; accepted
\ignorespaces
\the\ACCDATE\vskip21pt}}%
\edef\test{\the\THESAURUS}%
\ifx\test\missing\THESAURUS={missing; you have not inserted them}%
\message{|Thesaurus codes are not given.}\fi
\if M\REFEREE\let\REFEREE=Y
\normalbaselineskip=2\normalbaselineskip
\normallineskip=2\normallineskip\normalbaselines\fi
\edef\test{\the\SUMMARY}%
\ifx\test\missing\message{|Summary is missing.}%
\SUMMARY={Not yet given.}\fi
\noindent{\bf Summary. }\ignorespaces
\the\SUMMARY\vskip0.5true cm
\edef\test{\the\KEYWORDS}%
\ifx\test\missing\message{|Missing keywords.}%
\KEYWORDS={Not yet given.}\fi
\noindent{\bf Key words: }\the\KEYWORDS
\vskip3pt\line{\hrulefill}\vfill
\global\sterne=0
\catcode`\@=12}%


\hyphenation{have pow-er char-ac-teris-tic be-low Milne ve-loc-i-ty
 ver-i-fied
mag-net-ic ac-cel-era-tion be-tween rela-tion-ship col-lapse pro-vide
 that
con-sid-er-a-tion mas-sive found wind ap-prox-i-mate Astron be-yond}
\MAINTITLE={Cosmic Rays}
\SUBTITLE={I. The cosmic ray spectrum between $10^4 \; \rm GeV$ and
 $3
\;10^9\;
\rm GeV$}
\AUTHOR={\ts Peter L. Biermann}
\SENDOFF={\ts Peter L. Biermann}
\INSTITUTE={Max Planck Institut f\"ur Radioastronomie, Auf dem H\"ugel
 69,
D-5300 Bonn 1, Germany}
\RECDATE={August 1, 1992}
\ACCDATE={January 8, 1993}
\SUMMARY={ Based on a conjecture about the diffusion tensor of
 relativistic particles perpendicular to the magnetic field at a
shock, and considering particle drifts, I develop a theory to account
for the Cosmic Ray spectrum between $10^4 \; \rm GeV$ and $3 \;10^9 \;
\rm GeV$.  The essential assumption is that the free mean path
perpendicular to the magnetic field is independent of energy and has
the scale of the thickness of the shocked layer.  I then use the basic
concept, that the energetic Cosmic Ray particles areaccelerated in a
Supernova shock that travels down the density gradient of a stellar
wind; as an example I use a Wolf Rayet star wind. Physically important
ingredients beside the presence of a strong shock are diffusion,
drifts, convection, adiabatic cooling, the injection history, and the
topology of the magnetic field, assumed to behave similarly to the
solar wind.  The result is a spectrum, which for strong shocks and
negligible wind speeds in a gas with adiabatic index $5/3$ yields a
spectrum of $E^{-7/3}$.  Discussion of the latitude dependence of the
acceleration leads to a knee energy which is determined by an
expression of which the functional form leads to a suggestion on the
physical origin of the mechanical energy of Supernova explosions,
namely the gravitational potential energy mediated by the angular
momentum and the magnetic field.  Interstellar turbulence with a
Kolmogorov spectrum then leads by losses from the galactic disk to a
spectrum, which is $E^ {-8/3}$ below the knee, as observed in Cosmic
Rays, and as deduced from radio observations of the nonthermal
emission of our Galaxy as well as that of all other well observed
galaxies. At the knee the particles segregate with particle energy
according to their charge, with H dropping off first, then CNO
elements, then Mg, Si etc., and finally iron nuclei.  Further
consideration of the energy gain due to drifts at high particle
energies leads to a spectrum beyond the knee.  This spectrum is
$E^{-29/11}$ at injection, and, corrected for diffusive transport
through the Galaxy, very close to $E^{-3}$, as observed.  Beyond the
knee, iron and other heavy nuclei dominate out to the highest energies
of galactic Cosmic Ray particles. }

\KEYWORDS={Acceleration of particles -- Cosmic Rays -- Plasmas --
 Supernovae:
general -- Shockwaves}

\THESAURUS={02.01.1, 09.03.2, 02.16.1, 08.19.4, 02.19.1}

\maketitle

\titlea {Introduction}

The origin of Cosmic Rays is still not completely understood.  There
are few well accepted arguments: a) The Cosmic Rays below about $10^4
\; \rm GeV$ are predominantly due to the explosion of stars into the
normal interstellar medium (Lagage and Cesarsky 1983).  b) The Cosmic
Rays from near $10^4$ GeV up to the knee, at $5 \; 10^6 \; \rm GeV$
are very likely predominantly due to explosions of massive stars into
their former stellar wind (V\"olk and Biermann, 1988).  Clearly, there
is some overlap between the contributions from normal Supernova
explosions into the interstellar medium and explosions into a wind
cavity.  The consequences of this concept have been checked by
calculating the Cosmic Ray abundances and comparing them with
observations (Silberberg et al.  1990); the comparison suggests that
up to the highest energy where abundances are known, this concept
successfully explains the data. It is especially interesting, that no
direct mixing from the Supernova ejecta is required to account for the
known Cosmic Ray abundances, winds from red and blue supergiants as
well as Wolf Rayet winds as sources are all that is needed at present
in the high energy range below the knee.  c) For the energies beyond
the knee there is no consensus; Jokipii and Morfill (1987) argue that
a galactic wind termination shock might be able to provide those
particles, while Protheroe and Szabo (1992) argue for an extragalactic
origin, although in either case the matching of the flux at the knee
from two different source populations remains somewhat difficult.  d)
For the Cosmic Rays beyond the ankle at about $3 \; 10^9 \; \rm GeV$
an extragalactic origin is required because of the extremely large
gyroradii of such particles.

Biermann (1992), Rachen (1992), Rachen and Biermann (1992) have
proposed that these particles arise from hot spots in nearby radio
galaxies; this hypothesis leads to a successful and nearly
parameterfree explanation (Rachen 1992) of the intensity and spectrum
of these particles with the important proviso that the mean free path
in intergalactic space should be not much smaller than the
characteristic distances between the sources and us, and may be
similar to the scale of the large scale bubbles in the universe.

In all such arguments (also, e.g., Bogdan and V\"olk, 1983, Drury et
al.  1989, Markiewicz et al. 1990) the spectrum of the Cosmic Rays
remains largely unexplained.  And yet the observations of the Cosmic
Rays themselves, and of the nonthermal radioemission from our Galaxy
as well as from all other well observed galaxies (Golla 1989) strongly
suggests, that in all galactic environments studied carefully, the
Cosmic Rays have an universal spectrum of very nearly $E^{-8/3}$ below
the knee at $5 \,10^6$ GeV (electrons only $< 10$ GeV); direct air
shower experiments show the spectrum beyond the knee to be well
approximated by $E^{-3}$ (Stanev 1992).  This overall spectrum is
clearly influenced by propagation effects, since particles at
different energies have different chances to escape from the disk of
the galaxy.  It appears to be a reasonable hypothesis to approximate
the interstellar turbulence spectrum by a Kolmogorov law, which leads
to an interstellar diffusion coefficient proportional to $E^{1/3}$.
Such an energy dependence then requires a source spectrum of Cosmic
Rays of approximately $E^{-7/3}$ below the knee, and approximately
$E^{-8/3}$ above the knee.  In this paper I propose to derive such a
spectrum.

I note, that Ormes and Freier (1978) have argued already for such a
source spectrum.  The basic hypothesis is, again, that I consider
explosions into a stellar wind with a Parker spiral topology.  I
remind the reader that in such a wind in the asymptotic regime, where
the magnetic field decreases with the radius $r$ as $1/r$ and the wind
velocity is constant, the Alfv\'en velocity is also constant with
radius.

I also note that the wind in massive stars has a similar energy
integrated over the main sequence life time as the subsequent
supernova explosion.  The wind bubble is large and is itself
surrounded by a dense shell. Hence I can expect the shock of the
supernova to disperse this shell and to mix the energetic particle
population produced in the shock running through the wind directly
into the interstellar medium.  Thus, there is no additional energy
dependence introduced here to go from the spectrum which I calculate
below to the injection spectrum of cosmic rays.

In parallel and following papers I will test and explore the
consequences of the model proposed.

\titlea {The interstellar diffusive transport of Cosmic Rays}

There are arguments from the secondary to primary Cosmic Ray ratio
that at moderate energies the leakage time from the galactic disk is
about $10^7\; \rm years$ and has an energy dependence which varies as
$E^{0.6\pm0.1}$.  The higher energy data, however, suggest (Engelmann
et al. 1985, 1990) that the source spectrum is close to $E^{-2.4}$, so
that the leakage time varies with an energy exponent near $1/3$.  Data
from $1$ GeV/amu to $1$ TeV/amu (Swordy et al.  1990) suggest that the
leakage is down by one power of ten over these three decades in
particle energy, and so suggests again a leakage time energy
dependence at these energies of power $1/3$. The question of interest
in our context is how this time scale varies with high energy and how
its energy dependence can be related to the turbulence in the
interstellar medium.

The turbulence in the interstellar medium is known from pulsar
scintillation data and direct velocities of interstellar clouds to be
remarkably close to a Kolmogorov law (Larson 1979, 1981, Rickett
1990), which would translate to an $1/3$ power for the leakage time
energy dependence.  The high energy dependence of the diffusion
coefficient and the leakage time has to fulfill the condition that the
leakage time scale should not fall below the light travel time across
the disk because then, obviously, no transport is possible, and, in
addition, one should observe Cosmic Ray anisotropies.  Such
anisotropies are not observed and so, even at the highest energies of
those Cosmic Rays which I believe to come from sources in the galactic
disk, the leakage time has to be of order a few thousand years or
longer.  This condition is readily fulfilled by an energy dependence
of the leakage time of an exponent of $0.4$ or smaller, if the entire
energy range can be covered by one powerlaw.  On the other hand, if
the entire energy range is not covered by one single law, then there
should be a characteristic energy at which the energy dependence of
the diffusion changes; such a characteristic energy in turn
corresponds to an characteristic length: There is no evidence of any
special length scale in the interstellar medium between the
dissipation scale and the Larmor radius of thermal particles on the
one hand, and the thickness of the hot disk on the other hand, except
for the possibility of a characteristic length associated with giant
molecular clouds, as discussed further below.  Clearly, the simplest
hypothesis is to assume that the diffusive transport is governed by
basically one powerlaw; it is obvious that at low energies there are
likely to be variations on this due to source structure and cross
sections which almost certainly influence the secondaries energy
dependence.

Plasma simulations (Matthaeus and Zhou 1989) suggest that in a plasma,
where the energy density of the thermal gas and that of the magnetic
field are similar, the typical turbulence spectrum is indeed
Kolmogorov.  Thus basing my case on 1) observations of Cosmic Rays and
the interstellar medium as well as as on 2) theoretical work in plasma
physics which was developed for our physical understanding of the
solar wind, I assume that the leakage time has a $1/3$ powerlaw over
the entire energy range of interest.  Then the leakage time scale even
at $3 \; 10^9 \; \rm GeV$ is much longer than the light travel time
and so no anisotropy is expected.  With such a concept it becomes
obvious that I have to look for an explanation of a source spectrum of
Cosmic Rays which is approximately $E^{-7/3}$ below the knee and
$E^{-8/3}$ above the knee.

There is one scale in the interstellar medium which may be relevant:
The size of giant molecular clouds.  Giant molecular clouds are
assemblies of cloudlets that aggregate in a spiral arm.  With this
aggregation they can trap Cosmic Ray particles through the magnetic
fields, which permeate the clouds.  For particles with a mean free
path shorter than the size of the giant molecular clouds, trapping
occurs.  This limit defines a critical particle energy.  This trapping
leads to a production of secondaries (see also Dogiel and Sharov 1990)
which constitute an additional source term over and above that of the
normal interstellar medium. Here the secondaries have an injection
spectrum from the giant molecular cloud steeper than the normal
primaries by the energy dependence of the diffusion coefficient.
Assuming a Kolmogorov law also inside the cloud, which seems well
substantiated by observations (Larson 1981), this leads to a spectrum
steeper by $1/3$ than the standard Cosmic Ray spectrum.  This
population is then injected into the average interstellar medium upon
dissolution of the giant molecular cloud.  I emphasize that here the
time dependence of the process is important, since an equilibrium is
never established in this picture.  Since then at these low energies
these secondaries in the average interstellar medium are once more
subject to diffusive leakage from the disk, their spectrum is once
more steepened by another $1/3$, and so below the critical particle
energy defined above the secondaries are expected to show a spectrum
steeper by $2/3$ than the primaries.

\titlea {Definition of the task}

I consider the acceleration of particles in a shock that travels down
a steady stellar wind.  In this paper I will assume that the shock
speed is very much larger than the wind velocity; in Paper II
(Biermann and Cassinelli 1992), where those slower shocks in Wolf
Rayet and OB star winds are considered which cause nonthermal radio
emission through the acceleration of electrons, I will include the
effects of finite wind speeds.

The acceleration of particles is governed by the standard theory
(Parker 1965)

$$\eqalign{{\partial N \over \partial t}\;&=\;{ 1 \over r^2}\,{\partial
\over
{\partial r}}\, \left( r^2\, \kappa_{rr} \, {{\partial N} \over {\partial
 r}}
\right) \cr &+{1 \over r^2}\,{\partial \over {\partial \mu}}\,
\left(
(1-\mu^2)\,
\kappa_{\theta \theta}\,{{\partial N} \over {\partial \mu}}\,\right)
\cr
&-(V_r \,+\,V_{d,r})\,{{\partial N} \over {\partial r}} \cr
&+V_{d,\theta}\,(1-\mu^2)^{1/2}\,{1 \over r}\,{{\partial N} \over {
\partial
\mu}}
\cr &+{1 \over 3}\,{1 \over r^2}\,{{\partial V_r r^2} \over {
\partial
r}}\,{{\partial N} \over {\partial ln E}} \cr &+Q \cr} \eqno\autnum $$
where $N$ is the particle distribution function, and $Q$ is the
source; both are functions of the coordinates radial distance $r$,
colatitude $\theta$ or $\mu\,=\,\cos \,\theta$, time $t$ and particle
energy $E$, which I have taken to be relativistic.  The terms in the
above equation are first time change, then radial diffusion, latitude
diffusion, radial drift, latitude drift, then compression and finally
sources.  The two drift terms have a different sign, because I use the
cosine of the colatitude as my coordinate along $
\theta$.

The components of the diffusion tensor of interest here are the radial
diffusion term $\kappa_{rr}$ and the latitude diffusion term $\kappa_{
\theta\theta}$.  The drifts are also important with the radial component $V_
{d,r}$ and the latitude component $V_{d,\theta}$.  Outside the wind
acceleration region stellar winds are likely to be similar to the
solar wind, and so I will assume a Parker spiral topology of the
magnetic field (e.g. Jokipii et al.  1977):

$$(B_r,\,B_{\phi})\;=\;B_s\,\left( {r_s^2 \over r^2},\,-{r_s^2 \over
{r_W r}}\,(1-\mu^2)^{1/2} \right) . \eqno\autnum $$ Here $B_s$ is the
surface magnetic field of the star, assumed to be radial, $r_s$ is the
surface radius of the star, and $r_W\;=\;v_W/\Omega_s$ with $v_W$ and
$\Omega_s$ the wind velocity and the angular rotation rate of the
star.  I consider only distances $r$ much larger than either $r_s$ or
$r_W$.

This leads then to drift velocity components of

$$V_{d,r}\;=-\;{2 \over 3}\,c\,{E \over {Z e B_s}}\,{r_W^3 \over
{r_s^2 r^2}}\, {\mu \over {(1-\mu^2)^2}} \eqno\autnum $$ and

$$V_{d,\theta}\;=\;{2 \over 3}\,c\,{E \over {Z e B_s}}\,{r_W \over r_s^2}\,
{1
\over (1-\mu^2)^{1/2}} . \eqno\autnum $$ I note that, using the expression
 for the magnetic field above, this can also be written as

$$V_{d,\theta}\;=\;-{2 \over 3}\,c\,r_g/r, \eqno\autnum $$ where $r_g$
is the Larmor radius of the particle under consideration, and $r_g
\,< \,0$ and $V_{d,\theta} \, > \,0$, i.e. the drift is towards the
 equator, for $Z \,B_s \,> \, 0$ (see also Section 10).  This drift
here is just due to the unperturbed structure of the stellar wind
magnetic field; I will consider the consequences of additional
curvature from turbulence below.  At the equator I also have a radial
drift in the equatorial plane (Jokipii et al.  1977) which can be
written as

$$V_{d,r,equ}\;=\;-{2 \over 3}\,c\,\delta (\theta-\pi/2) \,r_g/r \eqno
\autnum
$$ with the same sign convention as in Eq. (5).  This drift will not
 become
important, however.

My boundary conditions are the usual: I inject particles at some low
particle energy which is assumed to be independent of all relevant
properties of the problem, i.e. not dependent on radial distance,
magnetic field strength or latitude.  The injection density depends,
of course, on the density of the wind medium.  Downstream I assume the
flow to take particles out of the system with the normal probability
$4\,U_2/c$ (see, e.g., Drury 1983), where $U_2$ is the downstream
velocity relative to the shock.  \par I propose to derive the
essential properties of the particle distribution function by analytic
means, using heuristic arguments.  Key will be the form of the
diffusion tensor, especially the radial component $\kappa_{rr}$.

\titlea {Conjecture: Diffusion perpendicular to the overall magnetic
 field}

I consider the propagation of a shock wave into a stellar wind which
has the standard Parker spiral magnetic-field structure. If the radial
diffusion coefficient increases linearly with $r$ (as assumed, for
example, by V\"olk and Biermann, see also, below) adiabatic loss time
and acceleration time in a first order Fermi theory have the same
radial dependence leading to the preservation of the highest energy
reached by particles until the shocks runs into the stellar wind
termination shell, well deep in the interstellar medium (V\"olk and
Biermann, 1988).  This can lead to much larger particle energies than
in alternative pictures, such as in an explosion into a homogeneous
interstellar medium in the Sedov expansion phase (Lagage and Cesarsky
1983).

Thus I am faced with the difficulty considering a shock which
propagates perpendicular to the magnetic field over almost all $4 \pi$
steradians.  In such a case particle drifts at the shock and in the
upstream and downstream regions are important (see, e.g., Jokipii
1987).

{\it Observations} can be a guide here: The explosions into the
interstellar medium are also explosions where the magnetic field is
nearly perpendicular to the shock direction over most of $4 \pi$
steradians.  Radio polarization observations of supernova remnants
yield clear evidence what the local structure of these shocked plasmas
typically is.  The observational evidence (Milne 1971, Downs and
Thompson 1972, Reynolds and Chevalier 1982, Milne 1987, Dickel et al.
1988) has been summarized by Dickel et al. (1991) in the statement
that all shell type supernova remnants less than $1000$ years age show
dominant radial structure in their magnetic fields near their
boundaries.  There are several possibilities to explain this:
Rayleigh-Taylor instabilities between ejected and swept up material
can lead to locally radial differential motion and so produce a
locally radial magnetic field (Gull 1973).  It could also be due to
strong radial velocity gradients of various ejecta, or due to overrun
clouds that now evaporate and cool the surrounding material.

The important conclusion for us here is that {\it observations
suggest} the existence of strong radial differential motions in
perpendicular shocks which in turn suggest that particles get {\it
convected} parallel to the shock direction.  I emphasize that
convective motion at a given scale entails that particle diffusion is
independent of energy.  I assume this convective turbulence with
associated particle transport to be a diffusive process, for which I
have to derive a natural velocity and a natural length scale, which
can be combined to yield a diffusion coefficient.  A classical
prescription is the method of Prandtl (1925) whose line of argument is
nicely reviewed and discussed by Stanisic (1988): In Prandtls argument
an analogy to kinetic gas theory is used to derive a diffusion
coefficient from a natural scale and a natural velocity of the system.
Despite many weaknesses of this generalization Prandtls theory has
held up remarkably well in many areas of physics far beyond the
original intent.  I will use a similar prescription here.  \par
Consider the structure of a layer shocked by a Supernova explosion
into a stellar wind in the case, that the adiabatic index of the gas
is $5/3$ and the shock is strong.  Then there is an inherent length
scale in the system, namely the thickness of the shocked layer, in the
spherical case for a shock velocity much larger than the wind speed
and in the strong shock limit $r/4$.  There is also a natural velocity
scale, namely the velocity difference of the flow with respect to the
two sides of the shock.  Both are the smallest dominant scale, in
velocity and in length; I will use the assumption that the smallest
dominant scale is the relevant scale several times in the course of
this paper in order to derive diffusion coefficients and other
scalings.

My basic conjecture, Postulate $1$, {\it based on observational
evidence}, is that the convective random walk of energetic particles
perpendicular to the magnetic field can be described by a diffusive
process with a downstream diffusion coefficient $\kappa_{rr,2}$ which
is given by the thickness of the shocked layer and the velocity
difference across the shock, and is independent of energy:

$$\kappa_{rr,2} \;=\;{1 \over 3}\, {U_2 \over U_1}\, r \, (U_1 \, -
\,U_2)
\eqno\autnum $$ The upstream diffusion coefficient can be derived in a
 similar way, but with a larger scale.  I make here the second
critical assumption, Postulate $2$, namely that the upstream length
scale is just $U_1/U_2$ times larger, and so is $r$.  This, obviously,
is the same ratio as the mass density and the ratio of the gyroradii
of the same particle energy.  Since the magnetic field is lower by a
factor of $U_1/U_2$ upstream, that means that the upstream gyroradius
of the maximum energy particle that could be contained in the shocked
layer, is also $r$.  Hence the natural scale is just $r$.  And so the
upstream diffusion coefficient is

$$\kappa_{rr,1} \;=\; {1 \over 3}\, r \, (U_1 \, - \, U_2)
\eqno\autnum $$ It immediately follows that the diffusive scales
relative to $r$ are

$${\kappa_{rr,1} \over {r \, U_1}}\;=\;{\kappa_{rr,2} \over {r \, U_2}}\;=
\;
{1
\over 3} \, (1 \, - \, {U_2 \over U_1}) \eqno\autnum $$

For these diffusion coefficients, it also follows that the residence
times (Drury 1983) on both sides of the shock are equal and are

$${4 \, \kappa_{rr,1} \over U_1 c}\;=\;{4 \,\kappa_{rr,2} \over U_2
c}\;=\; {4\over 3}\, {r \over c} \, (1 \, - \, {U_2 \over U_1}) .
\eqno\autnum $$

Adiabatic losses then cannot limit the energy reached by any particle
since they run directly with the acceleration time, both being
independent of energy, and so the limiting size of the shocked layer
limits the energy that can be reached to that where the gyroradius
just equals the thickness of the shocked layer, provided the particles
can reach this energy.  I assume here that the average of the magnetic
field $\langle B\rangle$ is not changed very much by all this
convective motion, but leave the possibility open that the root mean
square magnetic field ${\langle B^2
\rangle}^{1/2}$ is increased; this implies that a magnetic dynamo does
 not work as fast as the time scales given by the shock (see, e.g.,
Galloway and Proctor 1992 for arguments on the shortest possible
dynamo time scale).  This then leads to a maximum energy of

$$E_{max}\;=\; {U_2 \over U_1}\, Z e r B_2 \;=\; Z e r B_1
\eqno\autnum $$ where $Z e$ is the particle charge and $B_{1,2}$ is
the magnetic field strength on the two sides of the shock.  This
means, once again, that the energy reached corresponds to the maximum
gyroradius the system will allow on both sides of the shock.  It also
means that I push the diffusive picture right up its limit where on
the downstream side the diffusive scale becomes equal to the mean free
path and the gyroradius of the most energetic particles.

Jokipii (1987) has derived a general condition for possible values of
the diffusion coefficient: Its value has to be larger than the
gyroradius multiplied by the shock speed.  This condition is fulfilled
here, for the maximum energy particles only by a factor of $1-U_2/U_1
\, < \,1$, since here the shock speed and the radial scale of the
system give both the largest gyroradius as well as the diffusion
coefficient.

There is an important consequence of this picture for the diffusion
laterally:  From the residence timescale and the velocity difference
across the shock I find a distance which can be traversed in this time
of

$${4 \over 3}\, {r \over c} \, (1 \, - \, {U_2 \over U_1}) \, (U_1 \,-
\,U_2)
.  \eqno\autnum $$ Since the convective turbulence in the radial
direction also induces motion in the other two directions, with
maximum velocity differences of again $U_1 \,- \,U_2$, this distance
is also the the typical lateral length scale.   From this scale and
again the residence time I can construct an upper limit to the
diffusion coefficient in lateral directions of

$$\kappa_{\theta \theta ,max}\;=\;{4 \over 9} \, (1 \,-{U_2 \over U_1})^3
\,({U_1 \over c})^2 \, r \,c , \eqno\autnum $$ which is for strong
shocks equal to

$$\kappa_{\theta \theta ,max}\;=\;{1 \over 3} \, ({3 \over 4} \, {U_1
\over c})^2 \, r \,c .  \eqno\autnum $$

Again in the spirit of the idea, that the smallest dominant scale
wins, this then will begin to dominate as soon as the
$\theta$-diffusion coefficient reaches this maxium at a critical
energy.  As long as the $\theta$-diffusion coefficient is smaller, it
will dominate particle transport in $\theta$ and the upper limit
derived here is irrelevant.  When the $\theta$-diffusion coefficient
reaches and passes this maximum, then the particle in its drift will
no longer see an increased curvature due to the convective turbulence
due to averaging and the part ($1/3$ of total for strong shocks) of
drift acceleration due to increased curvature is eliminated.  This
then reduces the energy gain, and the spectrum becomes steeper from
that energy on.  The critical particle energy thus implied will be
identified below with the particle energy at the knee of the observed
cosmic ray spectrum.

\titlea {Particle Drifts}

Consider particles which are either upstream of the shock, or
downstream; as long as the gyrocenter is upstream I will consider the
particle to be there, and similarly downstream.

In general, the energy gain of the particles will be governed
primarily by their adiabatic motion in the electric and magnetic
fields.  The expression for the energy gain is then (Northrop, 1963,
equation 1.79), for an isotropic angular distribution

$${dE \over dt}\;=\;Z e {\bf V_d} {{{\bf U}\times {\bf B}} \over c}_i \,+
\,{pw \over c}{{\partial ln B} \over {\partial t}} \eqno\autnum $$ where
the first term arises from the drifts and the second from the induced
electric field.  This equation is valid in any coordinate frame. I
explicitly work in the shock frame, separate the two terms above and
consider the drift term first.  The second term is accounted for
further below, in Section 7.

The $\theta$-drift velocity in a normal stellar wind is given above,
in Section 3.  The $\theta$-drift can be understood as arising from
the asymmetric component of the diffusion tensor, the $\theta
r$-component.  The natural scales there are the gyroradius and the
speed of light, and so I note that for (Forman et al. 1974)

$$\kappa_{\theta r}\;=\;{1 \over 3} \,r_g \,c , \eqno\autnum $$ the
exact limiting form derived from ensemble averaging, I obtain the
drift velocity by taking the proper covariant divergence (Jokipii et
al. 1977); this is not simply (spherical coordinates) the
$r$-derivative of $\kappa_{\theta r}$.  The general drift velocity is
given by (see, e.g., Jokipii 1987)

$$V_{d,\theta}\;=\;c \,{E \over {3 Z e}}\,curl_{\theta} {{\bf B} \over B^2}
 .
\eqno\autnum $$ The $\theta$-drift velocity is thus:

$$V_{d,\theta}\;=\;{2 \over 3} \, c \, r_g/r \eqno\autnum $$ where
$r_g$ is now taken to be positive (see Eq. 5).  This drift velocity is
just that due to the gradient as well as the curvature, and in fact
both effects contribute here equally.  However, it must be remembered
that there is a lot of convective turbulence which increases the
curvature: The characteristic scale of the turbulence is $r/4$ for
strong shocks, and thus the curvature is $4/r$ maximum.  Taking half
the maximum as average I obtain then for the curvature a factor of
$2/r$ which is twice the curvature without any turbulence; this
increases the curvature term by a factor of two thus changing its
contribution from $1/3$ to $2/3$ in the numerical factor in the
expression above.  Hence the total drift velocity, combining now again
the curvature $(2/3)$ and gradient $(1/3)$ terms, is thus

$$V_{d,\theta}\;=\;{1 \over 3} \,(1 + {U_1 \over {2 U_2}})\, c \,
r_g/r ,\eqno\autnum $$ now written for arbitrary shock strength.  It
is easily verified that the factor in front is unity for strong shocks
where $U_1/U_2=4$.

The energy gain associated with such a drift is given by the product of
 the
drift velocity, the residence time, and the electric field.  Upstream
 this
energy gain is given by

$$\Delta E_1\;=\;{4 \over 3}\,E\,{U_1 \over c}\,f_d\, (1-{U_2 \over
 U_1}),
\eqno\autnum $$ where

$$f_d\;=\;{1 \over 3} \,(1 + {U_1 \over {2 U_2}}) . \eqno\autnum $$
Thus, $f_d=1$ for strong shocks. The corresponding expression
downstream is

$$\Delta E_2\;=\;=\;{4 \over 3}\,E\,{U_2 \over c}\,f_d\, (1-{U_2 \over
 U_1}) \eqno\autnum $$ giving a total energy gain of

$$\Delta E/E\;=\;{4 \over 3}\,{U_1 \over c}\,f_d\,(1+{U_2 \over
U_1})\,(1- {U_2
\over U_1}) \eqno\autnum $$ The drift energy gain averages over the
 magnetic field strength during the gyromotion.  I emphasize that this
energy gain is independent of this average magnetic field, so that
even variations of the magnetic field strength due to convective
motions do not change this energy gain.

It is of interest to note here, that the net distance travelled (i.e.
drifted) by the particle, e.g. upstream, is given by

$$l_{\perp 1}\;=\;{4 \over 3}\,{E \over {Z e B_1}} \,f_d\, (1 -{U_2
\over
U_1}) \eqno\autnum $$ which is the gyroradius itself for
$U_2/U_1=1/4$, corresponding to a strong shock.  This then says that
one is at a gyroradius limit for the drift distance, just as in
isotropic turbulence the gyroradius is a lower limit to the mean free
path for particle scattering parallel to the magnetic field in a
turbulent plasma, suggesting that it may be useful to think of the
plasma also as maximally turbulent perpendicular to the flow and
perpendicular to the magnetic field.  I emphasize that during this
drift the particle makes many gyromotions.  It is also important to
note that the magnetic field structure in the shocked region - as
discussed in Section 4 on the basis of observations - will contain
local regions of opposite magnetic field and so the drift itself will
be erratic and be the sum of many single element drift movements.
What I have derived is the average net energy gain due to drifts, with
the drift distance corresponding to the average magnetic field
strength.

\titlea {The energy gain of particles}

Let us consider then one full cycle of a particle remaining near the
shock and cycling back and forth from upstream to downstream and back.
The energy gain just due to the Lorentz transformations in one cycle
can then be written as

$${\Delta E \over E}_{LT}\;=\;{4 \over 3} {U_1 \over c} (1 -{U_2 \over
U_1}).\eqno\autnum $$ Adding the energy gain due to drifts I obtain

$${\Delta E \over E}\;=\;{4 \over3} {U_1 \over c} (1 -{U_2 \over U_1})
 x\eqno\autnum $$ where

$$x\;=\;1 + {1 \over 3}\,(1+{U_1 \over {2 U_2}})\,(1+{U_2 \over
 U_1})
\eqno\autnum $$ which is $9/4$ for a strong shock when $U_1 / U_2 \;=
\;4$.

Allowing for a general form of the diffusion coefficient, but keeping
here $\kappa_{rr,1}/U_1 \,=\,\kappa_{rr,2}/U_2$ for simplicity, I can
also write this result as

$$x\;=\;1\,+\,3\,{{\kappa_{rr,1}} \over {r U_1}}\,\tilde{f}_d \, (1+{U_2
\over
U_1})/(1-{U_2 \over U_1}) , \eqno\autnum $$ where now the generalized
factor $\tilde{f}_d$ is given by

$$\tilde{f}_d\;=\;{1 \over 3}\,(1 + {1 \over 6}\, {{r U_1} \over
{\kappa_{rr,1}}}\,{U_1 \over U_2}\,(1-{U_2 \over U_1}))
\eqno\autnum $$ This expression demonstrates how the effect of drifts
 gets smaller but does not go to zero with a smaller diffusion
coefficient.  On the other hand, obviously, given a geometry where the
magnetic field is parallel or nearly parallel to the shock normal, the
extra energy gain due to drifts no longer plays a role as, e.g., in
the polar cap.  \par It is easy to show that the additional energy
gain flattens the particle spectrum by

$${{3 \,U_2} \over {U_1 \, - \, U_2}} \,(1 \,- \,{1 \over x}) .\eqno\autnum
 $$

\titlea {Expansion and injection history}

Consider how long it takes a particle to reach a certain energy:

$${dt \over dE}\;=\;\lbrace 8 \;{\kappa_{rr,1} \over U_1 c} \rbrace /
\lbrace
{4
\over 3} {U_1 \over c} (1-{U_2 \over U_1}) x E \rbrace. \eqno\autnum $$ Here
 I
have used that $\kappa_{rr,1} / U_1 \;=\; \kappa_{rr,2} / U_2$. Since I
 have

$$r\;=\;U_1 t \eqno\autnum $$ this leads to

$${dt \over t}\;=\;{dE \over E} {{3 U_1} \over {U_1 - U_2}} {2 \over
 x}
{\kappa_{rr,1} \over {r U_1}} \eqno\autnum $$ and so to a dependence
 of

$$t(E)\;=\;t_o \; ({E \over E_o})^{\beta} \eqno\autnum $$ with

$$\beta\;=\;{{{3 U_1} \over {U_1 - U_2}} {2 \over x} {\kappa_{rr,1} \over
{r
U_1}}} \eqno\autnum $$ which is a constant independent of $r$ and
 $t$.

Particles that were injected some time ago were injected at a
different rate, say, proportional to $r^b$.  This then leads to a
correction factor for the abundance of

$$({E \over E_o})^{- b \beta}. \eqno\autnum $$ However, in a
$d$-dimensional space, particles have $r^d$ more space available to
them than when they were injected, and so I have another correction
factor which is

$$({E \over E_o})^{- d \beta}. \eqno\autnum $$ The combined effect is
a spectral change by

$$ -{{3 U_1} \over {U_1 - U_2}} {2 \over x} (d+b) {\kappa_{rr,1} \over
{r U_1}}.  \eqno\autnum $$ Thus I have a density correction factor,
which depends on the particle energy, and so changes the spectrum.
This expression can be compared with a limiting expansion derived by
Drury (1983; Eq. 3.58), who also allowed for a velocity field; Drury
(1983) generalized earlier work on spherical shocks by Krymskii and
Petukhov (1980) and Prishchep and Ptuskin (1981).  Drurys expression
agrees with the more generally derived expression given here for
$x=1$.  The comparison with Drurys work clarifies that for $\kappa
\,\sim \,r$ the inherent time dependence drops out except, obviously,
for the highest energy particles, discussed further below; the same
comparison shows that the statistics of the process are properly taken
into account in my simplified treatment. If the expansion is linear,
as is the case here, then the $r^d$-term also describes the adiabatic
losses in their effect on the spectrum, due to the general expansion
of the shock layer and thus accounts for the second term in Eq. (15)
above in Section 5.  Hence the total spectral difference, as compared
with the planeparallel case, is given by

$${{3 U_1} \over {U_1 - U_2}}\;\lbrace {U_2 \over U_1} ({1 \over x} -1) \;+
\;{2\over x} (b+d) {\kappa_{rr,1} \over {r U_1}} \rbrace. \eqno\autnum $$ Here
 I use the following sign convention: For this expression positive the
spectral index of the particle distribution is steeper than without
this correction; this then takes the minus sign in Eqs. (36 - 38)
properly into account.

This expression Eq. (39) together with Eq. (28) constitutes the basic
result of this paper; variants of this equation will be used below and
in subsequent communications for different modes or sites of
acceleration.  For a wind I have $b=-2$ and $d=+3$, and so
$b+d\;=\;1$.  The total spectral change is then for $U_1 / U_2 \;=\;4$
given by $1/3$, so that the spectrum obtained is

$${\rm Spectrum \,(source)}\;=\;E^{-7/3}. \eqno\autnum $$ This is what
I wanted to derive.  After correcting for leakage from the galaxy the
spectrum is

$${\rm Spectrum \,(earth)}\;=\;E^{-8/3} \eqno\autnum $$ very close to what
 is observed near earth at particle energies below the knee.

Such an injection spectrum of $-7/3$ of relativistic particles in
strong and fast shocks propagating through a stellar wind leads to an
unambiguous radio synchrotron emission spectrum of $\nu^{-2/3}$
(compare, e.g., the nonthermal radioemission of OB stars, WR stars,
novae, radio supernovae, and supernova remnants, which I plan to
discuss in subsequent communications).

It is of interest to note, that an injection spectrum of $-7/3$ can
also lead via pp-collisions in a synchrotron dominated regime to a
spectrum of pair-secondaries of $E^{-10/3}$, which translates in the
synchrotron spectrum to a $-7/6$ flux density spectrum and a $-13/6$
photon number spectrum, very close to that observed by GRO for the
Crab pulsar (Sch\"onfelder 1992, seminar in Bonn).  Such a speculative
interpretation would place the origin of the pulses in periodically
excited shocks travelling down a perpendicular magnetic field
configuration as considered here.

\titlea {The maximum energy of particles}

The maximum energy particles can reach is given in Section 4, and
depends linearly on the magnetic field.  Thus, I require estimates for
the magnetic field in the stellar winds of Wolf Rayet stars and other
massive stars, that explode as Supernovae, like red and blue
supergiants.  Comparing at first the corresponding estimates that
V\"olk and Biermann (1988) used, I note that the energies implied here
are larger by approximately $ c / U_1 $ for the same given magnetic
field strength, since their expression for the maximum energy that
particles could reach contains an additional factor of approximately
$U_1 / c$ as compared with my Eq. (11).

Cassinelli (1982), Maheswaran and Cassinelli (1988, 1992) have argued
that Wolf Rayet stars have very much larger magnetic fields than
V\"olk and Biermann used, in order to drive their winds.  The magnetic
fields given by Cassinelli and coworkers are of order a few thousand
Gauss on the surface of the star.  I introduce the conjecture here,
discussed in more detail in Paper II (Biermann and Cassinelli 1992),
that the Alfv\'en radius of the stellar wind is close to the stellar
surface itself.  Then it follows that the product $B r$ has
approximately the same value on the surface as in the wind, and is of
order $3\,10^{14} \, \rm cm \, Gauss$.  From this number I infer a
maximum energy of particles of

$$E_{max}({\rm protons})\;=\;9\,10^7 \, \rm GeV \eqno\autnum $$
 and

$$E_{max}({\rm iron})\;=\;3\,10^9 \, \rm GeV . \eqno\autnum $$ It
follows that the highest energy particles from the acceleration
process discussed here are mostly iron or other heavy nuclei.  The
chemical composition should change abruptly to mostly protons again
when the extragalactic component takes over (Rachen and Biermann 1992)
somewhere near $3\,10^9$ GeV.  If this mechanism provides the largest
particles energies, then obviously other contributions are not
excluded, by pulsars, neutron star binaries, or even from a
hypothetical termination of the galactic wind.

\titlea {The knee in the Cosmic Ray spectrum}

I wish to discuss here the bend in the spectrum of Cosmic Rays at the
knee, near $5 \; 10^6 \; \rm GeV$.

Let us consider the structure of the wind through which the supernova
shock is running. The maximum energy a particle can reach is
proportional to $\sin^2\theta$, since the space available for the
gyromotion from a particular latitude is limited in the direction of
the pole by the axis of symmetry.  Hence, clearly the maximum energy
attainable is lowest near the poles.  Then, consider the pole region
itself, where the radial dependence of the magnetic field is $1/r^2$,
and the magnetic field is mostly radial.  I can make two arguments
here: Either I put the upstream diffusive scale $4\;\kappa_ {rr,1} /(c
\,U_1)$ equal to $r/c$ in the strong shock limit, or I can put
acceleration time and flow time equal to each other.  Both arguments
lead to the same result.  Using here the Bohm limit in the diffusion
coefficient $\kappa_{rr,1}\;=\; {1 \over 3} c E/{Z e B(r)}$, since I
have a shock configuration near the pole, where the direction of
propagation of the shock is {\it parallel} to the magnetic field --
often referred to as a parallel shock configuration, then leads to a
maximum energy for the particles of

$$E\;=\;{3 \over 4} Z e B(r) r {U_1 \over c}, \eqno\autnum $$ which is
proportional to $1/r$ near the pole, where the magnetic field is
parallel to the direction of shock propagation; the corresponding
gyroradius is then given by ${3 \over 4} {U_1 \over c} r$.  Putting
this equal to the gyroradius of particles that are accelerated further
out at some colatitude $\theta$, where the magnetic field is nearly
perpendicular to the direction of shock propagation, gives the limit
where the latitude-dependent acceleration breaks down.  This then
gives the critical angle as

$$\sin \, \theta_{crit} \;=\; {3 \over 4} {U_1 \over c}. \eqno\autnum
$$ The angular range of $\theta \, < \, \theta_{crit}$ I refer to as
the {\it polar cap} below.  The energy at that location is then given
by

$$E_{knee}\; =\;Z e B(r) r ({3 \over 4} {U_1 \over c})^2. \eqno\autnum
$$ I identify this energy with the knee feature in the Cosmic Ray
spectrum, since all latitudes outside the polar cap contribute the
same spectrum up to this energy; from this energy to higher particle
energies a smaller part of the hemisphere contributes and also, the
energy gain is reduced, as argued below.  This is valid in the region
where the magnetic field is nearly perpendicular to the shock, and
thus this knee energy is independent of radius.

All this immediately implies that the chemical composition at the knee
changes so, that the gyroradius of the particles at the spectral break
is the same, implying that the different nuclei break off in order of
their charge $Z$, considered as particles of a certain energy (and not
as energy per nucleon).

In the polar cap the acceleration is a continuous mix between the
regime where the diffusion coefficient is determined by the thickness
of the shell, and the regime where it is dominated by turbulence
parallel to the magnetic field; this latter regime is rather small in
angular extent.  Thus, $\kappa_{rr,1}/ {r U_1}$ might be quite a bit
smaller than $1/4$.  Hence the polar cap will have a spectrum which is
determined by a range of

$$0\,<\,{\kappa_{rr,1} \over {r U_1}}\,<\,{1 \over 3}\,(1-{U_2 \over
 U_1}),
\eqno\autnum $$ as well as by a rather reduced role for the extra energy
 gain due to drifts.  This clearly corresponds (see Eqs. 28 and 39) to
a spectral index -- here again for simplicity in the strong shock
regime -- in the range $2$ to $7/3$.  Thus the spectrum of the
particles below the knee is likely to be flatter than $E^{-{7/3}}$ at
injection, or flatter than $E^{-{8/3}}$ after leakage from the galaxy.
The spectrum is harder in the polar cap region, because I am close to
the standard parallel shock configuration, for which the particle
spectrum is well approximated by $E^{-2}$.  On the other hand, the
polar cap is small relative to $4 \pi$ with about $(U_2 / U_1)^2$ and
only a spectrum much flatter than $E^{-7/3}$ like, e.g., indeed
$E^{-2}$ will make it possible for the polar cap to contribute
appreciably near the knee energy, because then the spectral flux near
the knee is increased relative to $1$ GeV by $(E_{knee} / m_p
c^2)^{1/3}$ which approximately compensates for its small area.
During an episode with drift towards the poles, a larger part of the
sphere can contribute for larger energy particles, and so there is an
additional tendency to flatten the spectrum of the polar cap
contribution.  The combination of the polar cap with the rest of the
stellar hemisphere might lead to a situation where up to, say, $10^4$
GeV the entire hemisphere excluding the polar cap dominates, while
from $10^4$ GeV up to the knee the polar cap begins to contribute
appreciably.  Near the knee energy the polar caps might thus
contribute equally to the rest of the $4 \pi$ steradians.  Because of
spatial limitations most of the hemisphere has to dominate again above
the knee, although with a fraction of the hemisphere that decreases
with particle energy. The superposition of such spectra for different
chemical elements will be tested elsewhere.  \par The expression for
the particle energy at the knee also implies by the observed relative
sharpness of the break of the spectrum that the actual values of the
combination $B(r) r U_1^2$ must be very nearly the same for all
supernovae that contribute appreciably in this energy range.  Please
note that $B(r) r$ is evaluated in the Parker regime, and so is
related to the surface magnetic field by $B(r) r \; =\; B_s
\; r_s^2 \Omega_s /v_W$, where the values with index $s$ refer to the
 surface of the star and $v_W$ is the wind velocity.  Thus the
expression

$$B_s \, r_s^2 \, {\Omega_s \over v_W} \, U_1^2 \eqno\autnum $$ is
approximately a universal constant for all stars that explode as
supernova after a Wolf Rayet phase.  It may also hold for all massive
stars of lower mass that explode as supernovae.

This then implies that I have found a functional relationship for the
mechanical energy of exploding stars connecting the magnetic field,
the angular momentum and the ejection energy.  Such a relationship
could be fortuitous, since all massive stars become very similar to
each other near the end of their evolution.  But it could also be an
indication for an underlying physical cause.  Related ideas have been
expressed and discussed by Kardashev (1970), Bisnovatyi-Kogan (1970),
LeBlanc and Wilson (1970), Ostriker and Gunn (1971), Amnuel et al.
(1972), Bisnovatyi-Kogan et al. (1976), and Kundt (1976), with
Bisnovatyi-Kogan (1970) the closest to the argument below.  All this
leads to the following interesting suggestion:

Consider a Wolf-Rayet star before is explodes as a supernova.  Given
my conjecture on the structure of the wind, i.e. that the Alfv\'en
radius is near the stellar surface, it is plausible to expect that WR
stars rotate not very far from critical at their surface.  On the
other hand, since WR stars represent the former inner convective cores
of OB stars, and in fact still have convective interiors, the magnetic
dynamo mechanism can be expected to operate and will produce maximum
magnetic fields of order $10^7$ Gauss for maximum rotation (see Paper
II).  Such magnetic fields make it plausible to assume that the stars
also rotate as a near-rigid body. The time scale of the transition
from a Wolf Rayet star to the final explosion is very short, and is
accompanied by the formation of further chemical abundance gradients
which slow any mixing. Hence it is plausible to assume that the
specific angular momentum distribution is not changed anymore in the
transition from the assumed solid body rotation in the Wolf Rayet
phase to the final explosion.  Using the models for Wolf-Rayet stars
of Langer (1989) with $Y\,=\, 1$, i.e. complete Helium composition,
the angular momentum $J_W$ of a Wolf Rayet star can then be written as

$$J_W \;=\;5.3 \; 10^{52} \,({M_W \over {5 \,M_{\odot}}})^{1.792}
\,j_W\,{\alpha}_W \, {\rm g \,cm^2 \,sec^{-1} } \eqno\autnum $$ where $j_W$ is
 the correction factor for the density structure which enters the
moment of inertia, and ${\alpha}_W$ is the fraction of critical
rotation at the surface, while $M_W$ is the mass of the Wolf rayet
star. I have scaled the properties from interpolating between the $5
\,M_{\odot}$ and $20 \, M_{\odot}$ mass models of Langer (1989)
representing the lower mass range for the most abundant Wolf Rayet
stars.  The angular momentum of that part of the stars which will form
a neutron star later, the innermost $1.4
\,M_{\odot}$ presumably, is given by

$$J_{ns,a}\;=\;6.7 \, 10^{50} \, ({M_W \over {5
\,M_{\odot}}})^{0.066}\,j_{ns,a}\,{\alpha}_W \, {\rm g \,cm^2 \,sec^
{-1}}
\eqno\autnum $$ where $j_{ns,a}$ is again a structural parameter for
 the density distribution which enters the moment of interia.  When
the star then implodes, the different mass shells halt their
contraction when they locally reach virial equilibrium between
rotation and gravitational energy.  The innermost region of the Wolf
Rayet star, that region destined to become a neutron star, collapses
most, but also reaches a halt in collapse due to virial equilibrium;
this is conceptually rather similar to the formation of a low mass
star out of a rotating cloud fragment - an accretion disk forms and
survives for a short time.  At this point the angular momentum is
conserved and so is

$$J_{ns,b}\;=\;J_{ns,a} . \eqno\autnum $$  From this condition one can
 derive
a radius $R_b$ of the central region which then determines the
 gravitational
energy available there.

$$R_b\;=\;3.1 \,10^8 \, ({ M_W \over {5\, M_{\odot}} })^{0.132}
\,({{j_{ns,a}\,{\alpha}_W } \over {j_{ns,b}} } )^2 \rm cm . \eqno\autnum
 $$ During collapse the magnetic field is irrelevant, since the
collapse time is much shorter than the time scale of angular momentum
transport by Alfv\'en waves.  However, when the collapse has halted in
the symmetry plane perpendicular to the rotational axis, torsional
Alfv\'en waves can transfer the rotational energy of the core to the
outer shells.  Because these outer shells are in virial equilibrium
before receiving this additional energy, the additional energy will
explode them.  The scale of the energy is easily shown to be of order
$10^{51}$ ergs for an initial state of rotation not far from breakup
at the surface:

$$E_b\;=\;1.7 \,10^{51} \,({ M_W \over {5\, M_{\odot}}} )^{-0.132}
\,({{j_{ns,a}\,{\alpha}_W } \over {j_{ns,b} } })^{-2} \rm erg . \eqno
\autnum$$ I note, that in this scenario the radius at which core collapse is
halted briefly, is of order $10^3$ the final neutron star radius, and
so densities are still comparatively low.  This radius is very close
to other estimates of the final precollapse stellar configurations,
and is here nearly independent of the initial Wolf Rayet stars mass;
the structural parameters $j_{ns,a}$ and $j_{ns,b}$ enter as ratios
and so also cancel to a large degree, and only the rotation parameter
$\alpha_W$ enters as a square, and hence the assumption of near
critical solid body rotation for the Wolf Rayet star is the most
important one in this argument (only relevant for the innermost part
of the star).  The final energy depends on the assumed mass of the
neutron star, here $1.4 \,M_{\odot}$, as $M_{ns}^{5/3}$.  \par In this
suggestion then, the source of the mechanical energy observable in
supernova explosions is then the gravitational energy at a scale
determined by angular momentum and mediated by the magnetic field.  I
leave a discussion how this argument may lead to a surface magnetic
field strength to later.

\titlea {The latitude distribution of the particles}

Consider Eq. (1) for the derivation of the spectrum beyond the knee.
Since the maximum energy particles can attain is a strong function of
colatitude, the spectrum beyond the knee requires a discussion of the
latitude distribution, which I have to derive first.  The latitude
distribution is established by the drift of particles which builds up
a gradient which in turn leads to diffusion down the gradient.  Hence
it is clear that drifts towards the equator lead to higher particle
densities near the equator, and drifts towards the poles lead to
higher particle densities there.  Thus the equilibrium latitude
distribution is given by the balancing of the $\theta$-diffusion and
the $\theta$-drift.

The diffusion tensor component $\kappa_{\theta \theta}$ can be derived
similar to my heuristic derivation of the radial diffusion term
$\kappa_{r r}$, again by using the smallest dominant scales.  The
characteristic velocity of particles in $\theta$ is given by the
erratic part of the drifting, corresponding to spatial elements of
different magnetic field direction, and this is on average the value
of the drift velocity $\mid V_{d,\theta}\mid$, possibly modified by
the locally increased values of the magnetic field strength, and the
characteristic distance is the distance to the symmetry axis
$r\,\sin\,\theta$; this is the smallest dominant scale as soon as the
thickness of the shocked layer is larger than the distance to the
symmetry axis, i.e. $\sin \,\theta\,<U_2 /U_1$.  Thus I can write in
this approximation, Postulate $3$,

$$\kappa_{\theta \theta ,1}\;=\;{1 \over 3}\,\mid V_{d,\theta}
\mid \,r\,(1-\mu^2)^{1/2} . \eqno\autnum $$ Here $\mu$ is again, see Eq. (1),
 the cosine of the colatitude on the sphere I consider for the shock
in the wind.  Interestingly, this can also be written in the form

$${1 \over 3}\,r_g \,c \, (1-\mu^2)^{1/2} , $$ where $r_g$ is taken as
positive; I also note that $c \, (1-\mu^2)^{1/2}$ is the maximum drift
speed at a given latitude, valid for the local maximum particle
energy.  This suggests that the latitude diffusion might be usefully
thought of as diffusion with a length scale of the gyroradius, and the
particle speed, to within the angular factor which just cancels out
the latitude dependence of the magnetic field strength in the
denominator of the gyroradius.

I assume then for the colatitude dependence a powerlaw
$(1-\mu^2)^{-a}$ and first match the latitude dependence of the
diffusion term and the drift, and then use the numerical coefficients
to determine the exponent in this law.  The diffusion term and the
drift term have the same colatitude dependence since the double
derivative and the internal factor of $(1-\mu^2)$ lead to a
$(1-\mu^2)^{-a-1}$ for the diffusive term, while the drift term is
just the simple derivative giving the same expression.  For
$(1-\mu^2)\,\ll\,1$ the condition then is

$$ {2 \over 3}\,a^2\;=\;\pm \,a.$$ It is important to remember the sign
 of
these terms.  The diffusive term is always positive, while the $
\theta$-drift term is negative for $Z \,B_s$ negative.  This means
for positive particles and a magnetic field directed inwards the
$\theta$-drift is towards the pole.  In that case then the exponent
$a$ is either zero or $a\,=\,3/2$. Since the drift itself clearly
produces a gradient, the case with $a\,=\,0$ is of no interest here.
It follows that for positive particles and an inwardly directed
magnetic field the latitude distribution is strongly biased towards
the poles, emphasizing in its integral the lower energies, and thus
making the overall spectrum steeper beyond the knee energy.  The
radial drift in this case is directed outwards, which means that
particles drift ahead of the shock by a small amount only to be caught
up again by the diffusive region ahead of the shock.  For the magnetic
field directed outwards and positive particles the radial drift is
inwards, taking particles out of the system at an slightly increased
rate and thus steepening the overall spectrum by a small amount.

When the magnetic field is directed outwards and the particles are
positive, the drift is towards the equator with then a positive
gradient with $(1-\mu^2)^{3/2}$, again in the limit
$(1-\mu^2)\,\ll\,1$.

I note that this exponent $3/2$ is reduced in the case, when the
erratic part of the drift is increased over the steady net drift
component.

\titlea {The range of the energy gain with drifting}

Consider for simplicity the case when the shock is strong, so that
$U_1/U_2\,=\,4$.

Ignoring at first the diffusive part of the drifting I find the
following: Then the combination of drift, acceleration and downstream
convection is given per cycle $\Delta n\,=\,1$

$$\Delta \mu\;=\;\pm\,{5 \over 4}\,{E \over E_{max}}\,\Delta n
\eqno\autnum $$ while

$${{\Delta E} \over E}\;=\;{3 \over 4}\,{U_1 \over c}\,\Delta n \eqno
\autnum $$ with the losses downstream given by

$$N\;=\;N_o\,(1-{U_1 \over c})^n . \eqno\autnum $$ This obviously is
constructed to give the overall spectrum derived earlier of power
$-7/3$.

The latitude drift can be integrated to give

$$\mu\;=\;\mu_o\,\pm\,(E/E_o-1)/b_E \eqno\autnum $$ with

$$b_E\;=\;{3 \over 5}\,{U_1 \over c}\,{E_{max} \over E_o} .
\eqno\autnum $$ It follows that even particles that drift down from
the region near the pole will not reach the maximum particle energy
possible in direct drifting, but are limited to

$$E_{\star}\;=\;{3 \over 5}\,{U_1 \over c}\,E_{max} , \eqno\autnum $$
which is formally very nearly the same as the maximum energy
attainable in the polar cap region, but {\it only} after allowing for
the different radial behaviour of the magnetic field there.  This
corresponds to the potential drop between the pole and the equator and
characterizes the anomalous component of the Cosmic Rays near the Sun.

However, the {\it diffusive part of the drifting}, responsible for the
$\theta$ diffusion, leads to additional energy gain - independent of
the strength of the magnetic field -, without a large net motion in
latitude, and so it is possible for particles to go to higher
energies.  The purely diffusive case leads in an analoguous
integration to a maximum energy of

$$E_{\star} \;=\;{4 \over 5} \,({3 \over 2} \,{U_1 \over c})^{1/2} \,E_{max}
 ,
\eqno\autnum $$ which is very close to $E_{max}$.  Hence the {\it
 combination
of direct and diffusive drifting} can be expected to yield particle
 energies
close to the maximum geometrically possible (Eq. 11).

\titlea {The spectrum beyond the knee}

This means that from $E_{knee}$ the energy gain of all particles in
the entire colatitude range that is affected by diffusion has to be
considered together.

In my model for the diffusion in $\theta$ I have used the drift
velocity and the distance to the symmetry axis as natural scales in
velocity and in length.  When the $\theta$-drift reaches the maximum
derived earlier, Eq. (13), then the latitude drift changes character.
This happens at a critical energy, which is reached at

$$\kappa_{\theta \theta,1}\;=\;\kappa_{\theta \theta , max} , \eqno\autnum
 $$ which translates into (see Eq. 46)

$$E_{crit} \;=\;({3 \over 4} \,{U_1 \over c})^2 \, E_{max} \;=\;
E_{knee} .\eqno\autnum $$ I emphasize that two different basically
geometric arguments lead to the same critical energy, $E_{knee}$.
This means, Postulate $4$, that for

$$E\;>\;E_{knee} \eqno\autnum $$ the drift energy gain is down by
$2/3$ to the level what the pure gradient and curvature drift yields,
Eq. (5), which results in

$$x\;=\;11/6 . \eqno\autnum $$ The reduced drift energy gain reduces
this value of $x$ below the limiting value derived earlier, of $9/4$.
This then leads to an overall spectrum of

$$E^{-29/11} , \eqno\autnum $$ before taking leakage into account, and

$$E^{-98/33}\;\cong\;E^{-3} , \eqno\autnum $$ with leakage accounted
 for. This is what I wanted to derive.

\titlea {Summary}

In the spirit of the Prandtl mixing length theory, often found to be
useful far beyond its original purpose, I make four postulates, all
defining components of the diffusion tensor for energetic particles
near perpendicular shocks in stellar winds:

Postulate 1 and 2 stipulate that the radial diffusion coefficient is
composed from the thickness of the shocked layer and the velocity
difference across the shock downstream and is larger by the ratio of
velocities upstream.  These diffusion tensor components are thus
independent of particle energy.  Using in addition the well
established connection between the transverse diffusion and drifts I
derive a) the spectrum of energetic particles below the knee.  Using
geometrical arguments I derive furthermore b) the maximum energy of
particles, c) the knee energy, d) the change of the chemical
composition of the energetic particle population at the knee, and e) a
suggestion of the physical origin of the mechanical energy of
supernova explosions of Wolf Rayet stars, possibly applicable to a
larger range of massive stars with winds.

Postulate 3 stipulates that the latitude diffusion coefficient is
composed from the drift speed and the distance to the symmetry axis
for particle energies below the knee energy, at which the drift speeds
reach the velocity difference across the shock at a critical latitude.
This leads to the latitude distribution of energetic particles (there
are two extreme cases depending on the orientation of the magnetic
field). For particle energies above the knee the latitude drift is
dominated by the lateral convective motions induced by the radial
convection.  This leads, Postulate $4$, to an elimination of any
contribution by increased curvature to the drift and thus, in the wind
case, to a reduction by a factor $2/3$ of the drift associated energy
gain.  From this postulate I derive f) the spectrum of energetic
particles beyond the knee, out to the maximum particle energies.

The relevant numbers all depend on adopting the high magnetic field
strengths proposed by Cassinelli et al. for Wolf Rayet stars; I
suggest that these high magnetic fields are common and argue that they
can be understood as the result of a dynamo working in the convection
zones in the interior of massive stars (for a detailed derivation of
the maxium magnetic field strength expected from dynamo theory
arguments, see Paper II).

I argue specifically about Wolf Rayet stars, but would like to suggest
that all massive stars with winds, red and blue supergiants, undergo a
similar fate, and thus may contribute the dominant part of the Cosmic
Ray spectrum from about $10^4$ GeV onwards, while at lower particle
energies explosions of massive stars also contribute.  But it is not
clear at present whether a) explosions into a homogeneous interstellar
medium, b) pulsar driven explosions or c) explosions into stellar
winds dominate at particle energies below $10^4$ GeV.  What I have
shown here is that the observed Cosmic Ray spectra also below $10^4$
GeV can be matched with the acceleration process in winds as described
here.  I plan to make the detailed comparison of these three sites for
particle acceleration in another communication.

I note that my approach relates the spectrum of accelerated particles
to the curvature of the shock, and thus, everything else being equal,
might help to disentangle -- or confuse -- inferences about the shock
from spectral indices observed in nonthermal radio emission.

It is clear that the analytic and heuristic arguments made here can
{\it only approximate the complexities} inherent in particle
acceleration in perpendicular shocks, in which the direction of shock
propagation is perpendicular to the magnetic field; on the other hand,
before a fully developed numerical treatment of the same processes
might become possible on highly parallel machines or with newly
developed sophisticated codes, I hope that analytic and semianalytic
progress to refine the theory and its concepts presented here will be
possible.

In this first paper of a series on the process of Cosmic Ray
acceleration I have introduced a basic conjecture (Section 4) on the
diffusion of particles in a shock perpendicular to the magnetic field.
In a similar vein I have introduced a number of heuristic arguments
which require {\it testing against observations}.  I plan to test all
implications of the model proposed here against available
observations.  I will further explore the consequences of this concept
in a following paper (with J.P. Cassinelli) on the nonthermal
radioemission of Wolf-Rayet stars and demonstrate that my concept can
produce the proper radio spectral indices, luminosities and temporal
behaviour.  In further communications I plan to apply this model to
explosions into the homogeneous interstellar medium, to the
radioemission of novae, and I will test in detail the predictions of
this model as regards the chemical abundances of Cosmic Rays.  What I
have shown here, is that the entire Cosmic Ray spectrum spectrum from
$10^4$ GeV out to $3\,10^9$ GeV, with the knee energy and the chemical
abundances, can be understood with the same physical ingredients
already well tested in the solar wind shock, namely a strong shock,
diffusion, drifts, convection, adiabatic cooling, injection history
and the familiar topology of the magnetic field in a realistic Parker
spiral.  The main requirements on the magnetic fields are that they as
strong as already implied by independent arguments from Wolf Rayet
star wind models, that help drive the wind with the rotating magnetic
field.

\ack {The essential part of this work was carried out during a
 5-month sabbatical in 1991 of at Steward Observatory at the
University of Arizona, Tucson.  I wish to thank Steward Observatory,
its director, Dr.  P.A.  Strittmatter, and all the local colleagues
for their generous hospitality during this time and during many other
visits.  Without a longstanding interaction and many discussions with
Dr. J.R. Jokipii this paper would not have started and would never
have been written; it is to him that my gratitude goes first.  I also
wish to thank Drs. D. Arnett, J.H. Bieging, J.P.  Cassinelli, D.
Eichler, T.K. Gaisser, G. Golla, N. Langer, K. Mannheim, W.H.
Matthaeus, H.  Meyer, R. Protheroe, M.M. Shapiro, T. Stanev and R.G.
Strom for extensive discussions of Cosmic Ray, Supernova and Star
physics. I wish to thank Drs.  T.K. Gaisser and W. Kr\"ulls for
comments on the manuscript.  High energy physics with PLB is supported
by grant DFG Bi 191/7 (Deutsche Forschungsgemeinschaft), BMFT grant
(DARA FKZ 50 OR 9202) and NATO travel grant CRG 910072. }

\begref
\ref Amnuel, P.R., Guseinov, O.Kh., Kasumov, F.K.: 1973 Sov. Astr.A.J. 16,
 932
(Astr.Zh. 49, 1139, 1972)
\ref Biermann, P.L.: 1992 in "Frontiers in Astrophysics", Eds.
 Silberberg,
R., Fazio, G., Rees, M., Cambridge Univ. Press (in press since
 1990)
\ref Biermann, P.L., Cassinelli, J.P.: 1992 (Paper II, in prep.)
\ref Bisnovatyi-Kogan, G.S.: 1970 Sov.Astr.A.J. 14, 652 (Astron.Zh. 47,
 813)
\ref Bisnovatyi-Kogan, G.S., Popov, Yu.P., Samochin, A.A.: 1976
 Astroph.Sp.Sc.
41, 287
\ref Bogdan, T.J., V\"olk, H.J.: 1983 Astron.\& Astroph. 122,
 129
\ref Cassinelli, J.P.: 1982 in "Wolf-Rayet stars: Observation,
 Physics,
Evolution", Eds. C.W.H. de Loore, A.J. Willis, Reidel, p. 173
\ref Dickel, J.R., Sault, R., Arendt, R.A., Matsui, Y., Korista, K.T.:
 1988
Astrophys.J. 330, 254
\ref Dickel, J.R., Breugel, W.J.M.van, Strom, R.G.: 1991 Astron.J. 101,
 2151
\ref Dogiel, V.A., Sharov, G.S.: 1990 Astron.\& Astrophys. 229,
 259
\ref Downs, G.S., Thompson, A.R.: 1972 Astron.J. 77, 120
\ref Drury, L.O'C: 1983 Rep.Prog.Phys. 46, 973
\ref Drury, L.O'C, Markiewicz, W.J., V\"olk, H.J.: 1989 Astron.
\&Astroph.
225, 179
\ref Engelmann, J.J., Goret, P., Juliusson, E., Koch-Miramond, L., Lund,
 N.,
Masse, P., Rasmussen, I.L., Soutoul, A.: 1985 Astron.\& Astroph. 148,
 12
\ref Engelmann, J.J., Ferrando, P., Soutoul, A., Goret, P., Juliusson,
 E.,
Koch-Miramond, L., Lund, N., Masse, P., Peters, B., Petrou, N.,
 Rasmussen,
I.L.: 1990 Astron.\& Astroph. 233, 96
\ref Forman, M.A., Jokipii, J.R., Owens, A.J.: 1974 Astrophys.J. 192,
 535
\ref Golla, G.: 1989 M.Sc. Thesis, University Bonn
\ref Galloway, D.J., Proctor, M.R.E.: 1992 Nature 356, 691
\ref Gull, S.F.: 1973 Monthly Not.Roy.Astr.Soc. 161, 47
\ref Jokipii, J.R., Levy, E.H., Hubbard, W.B.: 1977 Astrophys.J. 213,
 861
\ref Jokipii, J.R.: 1987 Astrophys.J. 313, 842
\ref Jokipii, J.R., Morfill, G.: 1987 Astrophys.J. 312, 170
\ref Kardashev, N.S.: 1970 Sov.Astr.A.J. 14, 375 (Astron.Zh. 47,
 465)
\ref Krymskii, G.F., Petukhov, S.I.: 1980 Sov.Astr.A.J.Letters 6, 124
 (Pis.
Astron.Zh. 6, 227)
\ref Kundt, W.: 1976 Nature 261, 673
\ref Lagage,P.O., Cesarsky, C.J.: 1983 Astron.\& Astroph. 118,
 223
\ref Langer, N.: 1989 Astron.\& Astroph. 210, 93
\ref Larson, R.B.: 1979 Monthly Not.Roy.Astr.Soc. 186, 479
\ref Larson, R.B.: 1981 Monthly Not.Roy.Astr.Soc. 194, 809
\ref LeBlanc, J.M., Wilson, J.R.: 1970 Astrophys.J. 161, 541
\ref Maheswaran, M., Cassinelli, J.P.: 1988 Astrophys.J. 335,
 931
\ref Maheswaran, M., Cassinelli, J.P.: 1992 Astrophys.J. 386,
 695
\ref Markiewicz, W.J., Drury, L.O'C., V\"olk, H.J.: 1990 Astron.
\&Astroph.
236, 487
\ref Matthaeus, W.H., Zhou, Y.: 1989 Phys.Fluids B 1, 1929
\ref Milne, D.K.: 1971 Austral.J.Phys. 24, 757
\ref Milne, D.K.: 1987 Austral.J.Phys. 40, 771
\ref Northrop, T. G.: 1963 "The Adiabatic Motion of Charged Particles",
Interscience Publishers, New York
\ref Ormes, J., Freier, P.: 1978 Astrophys.J. 222, 471
\ref Ostriker, J.P., Gunn, J.E.: 1971 Astrophys.J.Letters 164,
 L95
\ref Parker, E.N.: 1965 Planet. Space Sci.  13, 9
\ref Prandtl, L.: 1925 Zeitschrift angew. Math. und Mech. 5,
 136
\ref Prishchep, V.L., Ptuskin, V.S.: 1981 Sov.Astr.A.J. 25, 446
 (Astron.Zh.
58, 779)
\ref Protheroe, R.J., Szabo, A.P.: 1992 Preprint
\ref Rachen, J.: 1992 Master of Science Thesis, University of
 Bonn
\ref Rachen, J., Biermann, P.L.: 1992 in Proc. "Particle acceleration
 in
cosmic plasmas", Eds. G. Zank, T.K. Gaisser, AIP (in press)
\ref Reynolds, S.P., Chevalier, R.A.: 1982 Bull. American Astron. Soc. 14,
 936
\ref Rickett, B.J.: 1990 Ann.Rev.Astron.\&Astrophys. 28, 561
\ref Silberberg, R., Tsao, C.H., Shapiro, M.M., Biermann, P.L.:
 1990
Astrophys.J. 363, 265
\ref Stanev, T. 1992: In "Particle acceleration in cosmic plasmas", Eds.
 G.P.
Zank, T.K. Gaisser, AIP Conferenve Proc. No. 264, p. 379
\ref Stanisik, M.M.: 1988 "The Mathematical Theory of Turbulence",
 Springer
Verlag Berlin, 2nd edition
\ref V\"olk,H.J., Biermann, P.L.: 1988 Astrophys.J.Letters 333,
 L65
\ref Swordy, S.P., M\"uller, D., Meyer, P., L'Heureux, J., Grunsfeld,
 J.M.:
1990 Astrophys.J. 349, 625
\endref
\bye